\documentclass[12pt,letterpaper]{article}
\usepackage{epsfig,rotating,setspace,latexsym,amsmath,epsf,amssymb,bm,theorem}
\usepackage{cite,appendix}

\usepackage{amsfonts}

\title{Capacity Region of Gaussian MIMO Broadcast Channels with Common and Confidential Messages\thanks{This work was supported by NSF Grants CCF
04-47613, CCF 05-14846, CNS 07-16311 and CCF 07-29127.}}

\author{Ersen Ekrem \qquad Sennur Ulukus \\
\normalsize Department of Electrical and Computer Engineering\\
\normalsize University of Maryland, College Park, MD 20742 \\
\normalsize {\it ersen@umd.edu} \qquad {\it ulukus@umd.edu}}

\newcommand{\bblambda}{\bm \Lambda}

\newcommand{\bbsigma}{\bm \Sigma}

\newcommand{\bbi}{{\mathbf{I}}}
\newcommand{\bzero}{{\mathbf{0}}}
\newcommand{\bbv}{{\mathbf{V}}}

\newcommand{\bbh}{{\mathbf{H}}}

\newcommand{\bbm}{{\mathbf{M}}}

\newcommand{\bbk}{{\mathbf{K}}}

\newcommand{\bbn}{{\mathbf{N}}}

\newcommand{\bba}{{\mathbf{A}}}

\newcommand{\bbb}{{\mathbf{B}}}

\newcommand{\bbs}{{\mathbf{S}}}

\newcommand{\bbu}{{\mathbf{U}}}

\newcommand{\bbx}{{\mathbf{X}}}

\newcommand{\bby}{{\mathbf{Y}}}

\newtheorem{Theo}{Theorem}

\newtheorem{Lem}{Lemma}
\newtheorem{Cor}{Corollary}

\begin{document}


\maketitle

\begin{abstract}
We study the two-user Gaussian multiple-input multiple-output
(MIMO) broadcast channel with common and confidential messages. In
this channel, the transmitter sends a common message to both
users, and a confidential message to each user which needs to be
kept perfectly secret from the other user. We obtain the entire
capacity region of this channel. We also explore the connections
between the capacity region we obtain for the Gaussian MIMO
broadcast channel with common and confidential messages and the
capacity region of its non-confidential counterpart, i.e., the
Gaussian MIMO broadcast channel with common and private messages,
which is not known completely.

\end{abstract}

\newpage

\section{Introduction}
We consider the two-user Gaussian multiple-input multiple-output
(MIMO) broadcast channel, where each link between the transmitter
and each user is modelled by a linear additive Gaussian channel.
We study the two-user Gaussian MIMO broadcast channel for the
following scenario: The transmitter sends a common message to both
users, and a confidential message to each user which needs to be
kept perfectly secret from the other user. We call the channel
model arising from this scenario the {\it Gaussian MIMO broadcast
channel with common and confidential messages}.

The Gaussian MIMO broadcast channel with common and confidential
messages subsumes several other channel models as special cases.
The first one is the Gaussian MIMO wiretap channel, where the
transmitter has only one confidential message for one (legitimate)
user, which is kept perfectly secret from the other user
(eavesdropper). The secrecy capacity of the Gaussian MIMO wiretap
channel is obtained in \cite{Hassibi,Wornell} for the general
case, in \cite{Ulukus} for the 2-2-1 case. The second channel
model that the Gaussian MIMO broadcast channel with common and
confidential messages subsumes is the Gaussian MIMO wiretap
channel with common message~\cite{Liu_Common_Confidential}, in
which the transmitter sends a common message to both the
legitimate user and the eavesdropper, and a confidential message
to the legitimate user that is kept perfectly secret from the
eavesdropper. The capacity region of the Gaussian MIMO wiretap
channel with common message is obtained
in~\cite{Liu_Common_Confidential}. The third channel model that
the Gaussian MIMO broadcast channel with common and confidential
messages encompasses is the Gaussian MIMO broadcast channel with
confidential messages~\cite{Ruoheng_MIMO_BC}, where the
transmitter sends a confidential message to each user which is
kept perfectly secret from the other user. The capacity region of
the Gaussian MIMO broadcast channel with confidential messages is
established in~\cite{Ruoheng_MIMO_BC}.

Here, we obtain the capacity region of the Gaussian MIMO broadcast
channel with common and confidential messages. In particular, we
show that a variant of the secret dirty-paper coding (S-DPC)
scheme proposed in~\cite{Ruoheng_MIMO_BC} is capacity-achieving.
Since the S-DPC scheme proposed in~\cite{Ruoheng_MIMO_BC} is for
the transmission of only two confidential messages, it is modified
here to incorporate the transmission of a common message as well.
Similar to~\cite{Ruoheng_MIMO_BC}, we also notice an invariance
property of this achievable scheme with respect to the encoding
order used in the S-DPC scheme. In other words, two achievable
rate regions arising from two possible encoding orders used in the
S-DPC scheme are identical, and equal to the capacity region. We
provide the proof of this statement as well as the converse proof
for the capacity region of the Gaussian MIMO broadcast channel
with common and confidential messages by using the channel
enhancement technique~\cite{Shamai_MIMO} and an extremal
inequality from~\cite{Liu_Compound}.

We also explore the connections between the Gaussian MIMO
broadcast channel with common and confidential messages and its
non-confidential counterpart, i.e., the (two-user) Gaussian MIMO
broadcast channel with common and private messages. In the
Gaussian MIMO broadcast channel with common and private messages,
the transmitter again sends a common message to both users, and a
private message to each user, for which there is no secrecy
constraint now, i.e., private message of each user does not need
to be kept secret from the other user. Thus, the Gaussian MIMO
broadcast channel with common and confidential messages we study
here can be viewed as a constrained version of the Gaussian MIMO
broadcast channel with common and private messages, where the
constraint comes through forcing the private messages to be
confidential. We note that although there are partial results for
the Gaussian MIMO broadcast channel with common and private
messages~\cite{Hannan_Common,hannan_thesis}, its capacity region
is not known completely. However, here, we are able to obtain the
entire capacity region for a constrained version of the Gaussian
MIMO broadcast channel with common and private messages, i.e., for
the Gaussian MIMO broadcast channel with common and confidential
messages. We provide an intuitive explanation of this
at-first-sight surprising point as well as the invariance property
of the achievable rate region with respect to the encoding orders
that can be used in the S-DPC scheme, by using a result
from~\cite{hannan_thesis} for the Gaussian MIMO broadcast channel
with common and private messages. In particular, we use the
following result from~\cite{hannan_thesis}: For a given common
message rate, the private message sum rate capacity of the
Gaussian MIMO broadcast channel with common and private messages
is achieved by the dirty-paper coding (DPC) scheme
in~\cite{Goldsmith_common}, and any one of the two possible
encoding orders that can be used in DPC gives the private message
sum rate capacity. Using this result, we show that there is a
one-to-one correspondence between the points on the boundary of
the achievable rate region of the Gaussian MIMO broadcast channel
with common and confidential messages that are obtained by using a
specific encoding order in the S-DPC scheme, and those points
which are private message sum rate capacity achieving for the
Gaussian MIMO broadcast channel with common and private messages.
This correspondence intuitively explains why the achievable rate
regions arising from the use of different encoding orders in S-DPC
are the same, and also why we can obtain the entire capacity
region of the Gaussian MIMO broadcast channel with common and
confidential messages although the capacity region of its
non-confidential counterpart is not known completely.

\section{Channel Model and Main Result}
We study the two-user Gaussian MIMO broadcast channel which is
defined by
\begin{align}
\bby_1&=\bbh_1 \bbx+\bbn_1 \label{general_gaussian_mimo_1}\\
\bby_2&=\bbh_2 \bbx+\bbn_2 \label{general_gaussian_mimo_2}
\end{align}
where the channel input $\bbx$ is a $t\times 1$ vector, $\bbh_j$
is the channel gain matrix of size $r_j\times t,$ the channel
output of the $j$th user $\bby_j$ is a $r_j\times 1$ vector, and
the Gaussian random vector $\bbn_j$ is of size $r_j\times 1$ with
a covariance matrix $\bbsigma_j$ which is assumed to be strictly
positive-definite, i.e., $\bbsigma_j\succ \bzero$. We consider a
covariance constraint on the channel input as follows
\begin{align}
E\left[\bbx\bbx^{\top}\right] \preceq \bbs
\label{covariance_constraint}
\end{align}
where $\bbs \succeq \bzero$.

We study the following scenario for the Gaussian MIMO broadcast
channel: There are three independent messages $(W_0,W_1,W_2)$ with
rates $(R_0,R_1,R_2)$, respectively, where $W_0$ is the common
message that needs to be delivered to both users, $W_1$ is the
confidential message of the first user which needs to be kept
perfectly secret from the second user, and similarly, $W_2$ is the
confidential message of the second user which needs to be kept
perfectly secret from the first user. The secrecy of the
confidential messages is measured by the normalized equivocation
rates~\cite{Wyner,Korner}, i.e, we require
\begin{align}
\frac{1}{n}I(W_1;W_0,W_2,\bby_{2}^n)\rightarrow 0~~{\rm and}~~
\frac{1}{n}I(W_2;W_0,W_1,\bby_{1}^n)\rightarrow 0
\label{equivocation_1}
\end{align}
as $n\rightarrow \infty$, where $n$ denotes the number of channel
uses. The closure of all achievable rate triples $(R_0,R_1,R_2)$
is defined to be the capacity region, and will be denoted by
$\mathcal{C}(\bbs)$. We next define the following shorthand
notations
\begin{align}
R_{0j}(\bbk_1,\bbk_2)&=\frac{1}{2} \log\frac{|\bbh_j \bbs
\bbh_j^{\top}+\bbsigma_j|}{|\bbh_j(\bbk_1+\bbk_2)\bbh_j^\top+\bbsigma_j|},\quad j=1,2 \label{common}\\
R_1(\bbk_1,\bbk_2)&=\frac{1}{2}\log\frac{|\bbh_1(\bbk_1+\bbk_2)\bbh_1^\top+\bbsigma_1|}{|\bbh_1
\bbk_2\bbh_1^\top+\bbsigma_1|}
-\frac{1}{2}\log\frac{|\bbh_2(\bbk_1+\bbk_2)\bbh_2^\top+\bbsigma_2|}{|\bbh_2 \bbk_2\bbh_2^\top+\bbsigma_2|}\label{conf_1}\\
R_2(\bbk_2)&=
\frac{1}{2}\log\frac{|\bbh_2\bbk_2\bbh_2^\top+\bbsigma_2|}{|\bbsigma_2|}-\frac{1}{2}\log\frac{|\bbh_1\bbk_2\bbh_1^\top+\bbsigma_1|}{|\bbsigma_1|}
\label{conf_2}
\end{align}
using which, our main result can be stated as follows.
\begin{Theo}
\label{theorem_main_result} The capacity region of the Gaussian
MIMO broadcast channel with common and confidential messages
$\mathcal{C}(\bbs)$ is given by
\begin{align}
\mathcal{C}(\bbs)=\mathcal{R}_{12}^{\rm
S-DPC}(\bbs)=\mathcal{R}_{21}^{\rm S-DPC} (\bbs)
\end{align}
where $\mathcal{R}_{12}^{\rm S-DPC}(\bbs)$ is given by the union
of rate triples $(R_0,R_1,R_2)$ satisfying
\begin{align}
R_0&\leq \min\{R_{01}(\bbk_1,\bbk_2),R_{02}(\bbk_1,\bbk_2)\} \\
R_1&\leq R_1(\bbk_1,\bbk_2)\\
R_2&\leq R_2(\bbk_2)
\end{align}
for some positive semi-definite matrices $\bbk_1,\bbk_2$ such that
$\bbk_1+\bbk_2\preceq \bbs$, and $\mathcal{R}_{21}^{\rm
S-DPC}(\bbs)$ can be obtained from $\mathcal{R}_{12}^{\rm
S-DPC}(\bbs)$ by swapping the subscripts $1$ and 2.
\end{Theo}

Theorem~\ref{theorem_main_result} states that the common message,
for which a covariance matrix $\bbs-\bbk_1-\bbk_2$ is allotted,
should be encoded by using a standard Gaussian codebook, and the
confidential messages, for which covariance matrices
$\bbk_1,\bbk_2$ are allotted, need to be encoded by using the
S-DPC scheme proposed in~\cite{Ruoheng_MIMO_BC}. S-DPC is a
modified version of DPC~\cite{Wei_Yu} to meet the secrecy
requirements. The receivers first decode the common message by
treating the confidential messages as noise, and then each
receiver decodes the confidential message intended to itself.
Depending on the encoding order used in S-DPC, one of the users
gets a clean link for the transmission of its confidential
message, where there is no interference originating from the other
user's confidential message. Although one might expect that the
two achievable regions arising from two possible encoding orders
that can be used in S-DPC could be different, i.e.,
$\mathcal{R}_{12}^{\rm S-DPC}(\bbs)\neq \mathcal{R}_{21}^{\rm
S-DPC}(\bbs)$, and taking a convex closure of these two regions
would yield a larger achievable rate region,
Theorem~\ref{theorem_main_result} states that
$\mathcal{R}_{12}^{\rm S-DPC}(\bbs)= \mathcal{R}_{21}^{\rm
S-DPC}(\bbs)$, i.e., the achievable rate region is invariant with
respect to the encoding order used in S-DPC. This invariance
property of S-DPC was first noticed in~\cite{Ruoheng_MIMO_BC} for
the case where there was no common message to be transmitted.

\subsection{Aligned Channel}
We define a sub-class of Gaussian MIMO broadcast channels called
the aligned Gaussian MIMO broadcast channel, which can be obtained
from
(\ref{general_gaussian_mimo_1})-(\ref{general_gaussian_mimo_2}) by
setting $\bbh_1=\bbh_2=\bbi$, i.e.,
\begin{align}
\bby_1&=\bbx+\bbn_1\label{aligned_channel_1}\\
\bby_2&=\bbx+\bbn_2 \label{aligned_channel_2}
\end{align}
To distinguish the notation used for the aligned Gaussian MIMO
broadcast channel from the one used for the general model in
(\ref{general_gaussian_mimo_1})-(\ref{general_gaussian_mimo_2}),
we denote the capacity region of the aligned channel by
$\mathcal{C}^{\rm AL}(\bbs)$, the rate expressions in
(\ref{common})-(\ref{conf_2}) for the special case
$\bbh_1=\bbh_2=\bbi$ by $ \{R_{0j}^{\rm
AL}(\bbk_1,\bbk_2)\}_{j=1}^2,R_1^{\rm AL}(\bbk_1,\bbk_2),R_2^{\rm
AL}(\bbk_2)$, and the regions $\mathcal{R}_{12}^{\rm
S-DPC}(\bbs),\mathcal{R}_{21}^{\rm S-DPC}(\bbs)$ for the special
case $\bbh_1=\bbh_2=\bbi$ by $\mathcal{R}_{12}^{\rm
S-DPC-AL}(\bbs),\mathcal{R}_{21}^{\rm S-DPC-AL}(\bbs)$.

In this work, we first prove Theorem~\ref{theorem_main_result} for
the aligned Gaussian MIMO broadcast channel. Then, we establish
the capacity region for the general channel model in
(\ref{general_gaussian_mimo_1})-(\ref{general_gaussian_mimo_2}) by
following the analysis in Section~V.B of~\cite{Shamai_MIMO} and
Section~7.1 of~\cite{MIMO_BC_Secrecy} in conjunction with the
capacity result we obtain for the aligned channel.

\subsection{Capacity Region under a Power Constraint}

We note that the covariance constraint on the channel input in
(\ref{covariance_constraint}) is a rather general constraint that
subsumes the power constraint
\begin{align}
E\left[\bbx^{\top}\bbx\right]={\rm tr}\left(E\left[\bbx
\bbx^{\top}\right]\right)\leq P \label{trace_constraint}
\end{align}
as a special case, see Lemma~1 and Corollary~1
of~\cite{Shamai_MIMO}. Therefore, using
Theorem~\ref{theorem_main_result}, the capacity region arising
from the average power constraint in (\ref{trace_constraint}),
$\mathcal{C}(P)$, can be found as follows.
\begin{Cor}
The capacity region of the Gaussian MIMO broadcast channel with
common and confidential messages subject to a power constraint
$P$, $\mathcal{C}(P)$, is given by
\begin{align}
\mathcal{C}(P)=\mathcal{R}_{12}^{\rm
S-DPC}(P)=\mathcal{R}_{21}^{\rm S-DPC} (P)
\end{align}
where $\mathcal{R}_{12}^{\rm S-DPC}(P)$ is given by the union of
rate triples $(R_0,R_1,R_2)$ satisfying
\begin{align}
R_0&\leq \min\{R_{01}(\bbk_1,\bbk_2,\bbk_c),R_{02}(\bbk_1,\bbk_2,\bbk_c)\} \\
R_1&\leq R_1(\bbk_1,\bbk_2)\\
R_2&\leq R_2(\bbk_2)
\end{align}
for some positive semi-definite matrices $\bbk_1,\bbk_2,\bbk_c$
such that ${\rm tr}(\bbk_1+\bbk_2+\bbk_c)\leq P$, and
$\{R_{0j}(\bbk_1,\bbk_2,\bbk_c)\}_{j=1}^2$ are defined as
\begin{align}
R_{0j}(\bbk_1,\bbk_2,\bbk_c)&=\frac{1}{2} \log\frac{|\bbh_j
(\bbk_1+\bbk_2+\bbk_c)
\bbh_j^{\top}+\bbsigma_j|}{|\bbh_j(\bbk_1+\bbk_2)\bbh_j^\top+\bbsigma_j|},\quad
j=1,2
\end{align}
Moreover, $\mathcal{R}_{21}^{\rm S-DPC}(P)$ can be obtained from
$\mathcal{R}_{12}^{\rm S-DPC}(P)$ by swapping the subscripts $1$
and 2.
\end{Cor}
\vspace{-0.25cm}
\section{Proof of Theorem~\ref{theorem_main_result} for the Aligned Case}

\vspace{-0.25cm}
\subsection{Achievability}
\label{sec:ach_aligned} Here, we prove the achievability of the
regions $\mathcal{R}_{12}^{\rm S-DPC-AL}(\bbs)$ and
$\mathcal{R}_{21}^{\rm S-DPC-AL}(\bbs)$. To this end, we consider
the two-user discrete memoryless channel for the scenario where a
common message is delivered to both users, and each user gets a
confidential message which needs be kept perfectly secret from the
other user. For this scenario, we have the following achievable
rate region~\cite{Biao_Chen}.
\begin{Lem}[\!\!\cite{Biao_Chen}, Theorem~1]
\vspace{-0.35cm} \label{lemma_ach} The rate triples
$(R_0,R_1,R_2)$ satisfying
\begin{align}
R_0&\leq \min\{I(U;Y_1),I(U;Y_2)\}\\
R_1&\leq \left[I(V_1;Y_1|U)-I(V_1;Y_2,V_2|U)\right]^+ \\
R_2&\leq \left[I(V_2;Y_2|U)-I(V_2;Y_1,V_1|U)\right]^+
\end{align}
for some $(U,V_1,V_2)$ such that $(U,V_1,V_2)\rightarrow
X\rightarrow (Y_1,Y_2)$\footnote{In~\cite{Biao_Chen}, the
necessary Markov chain that $(U,V_1,V_2,X,Y_1,Y_2)$ needs to
satisfy is given by $U\rightarrow (V_1,V_2)\rightarrow X
\rightarrow (Y_1,Y_2)$. However, their achievable rate region is
valid for the looser Markov chain $(U,V_1,V_2)\rightarrow
X\rightarrow (Y_1,Y_2)$ as well, which we use here.} are
achievable.
\end{Lem}
We now use Lemma~\ref{lemma_ach} to show the achievability of the
region $\mathcal{R}^{\rm S-DPC-AL}_{12}(\bbs)$. We first introduce
three independent Gaussian random vectors $\bbu_0,\bbu_1,\bbu_2$
with covariance matrices $\bbs-\bbk_1-\bbk_2,\bbk_1,\bbk_2$,
respectively. Using these Gaussian random vectors, we set the
auxiliary random variables in Lemma~\ref{lemma_ach} as follows
\begin{align}
U&=\bbu_0 \label{auxiliary_U} \\
V_1&=\bbu_1+\bbu_0 \label{auxiliary_V1}\\
V_2&=\bbu_2+\bba\bbu_1+\bbu_0 \label{auxiliary_V2}
\end{align}
where $\bba=\bbk_2\left[\bbk_2+\bbsigma_2\right]^{-1}$ is the
precoding matrix for the second user to suppress the interference
originating from $\bbu_1$~\cite{Wei_Yu}. Furthermore, we set the
channel input $\bbx$ as follows
\begin{align}
\bbx=\bbu_0+\bbu_1+\bbu_2 \label{channel_input}
\end{align}
Using the definitions in (\ref{auxiliary_U})-(\ref{channel_input})
for the common message rate given in Lemma~\ref{lemma_ach}, we get
\begin{align}
R_0&=\min\left\{\frac{1}{2}\log
\frac{|\bbs+\bbsigma_1|}{|\bbk_1+\bbk_2+\bbsigma_1|},\frac{1}{2}\log\frac{|\bbs+\bbsigma_2|}{|\bbk_1+\bbk_2+\bbsigma_2|}\right\}
\end{align}
We next compute the second user's confidential message rate as
follows
\begin{align}
R_2&=I(V_2;\bby_2|U)-I(V_2;\bby_1,V_1|U)\\
&=I(\bbu_2+\bba\bbu_1;\bbu_1+\bbu_2+\bbn_2)-I(\bbu_2+\bba
\bbu_1;\bbu_1+\bbu_2+\bbn_1,\bbu_1)\\
&=I(\bbu_2+\bba\bbu_1;\bbu_1+\bbu_2+\bbn_2)-I(\bbu_2+\bba
\bbu_1;\bbu_1)-I(\bbu_2;\bbu_2+\bbn_1)\\
&=\frac{1}{2}
\log\frac{|\bbk_2+\bbsigma_2|}{|\bbsigma_2|}-I(\bbu_2;\bbu_2+\bbn_1)\label{wei_yu_implies}\\
&=\frac{1}{2}
\log\frac{|\bbk_2+\bbsigma_2|}{|\bbsigma_2|}-\frac{1}{2}
\log\frac{|\bbk_2+\bbsigma_1|}{|\bbsigma_1|}
\end{align}
where (\ref{wei_yu_implies}) is due to Theorem~1 in~\cite{Wei_Yu}.
We next compute the first user's confidential message rate. To
this end, we note the following
\begin{align}
I(V_1;\bby_2,V_2|U)&=I(V_1;\bby_2|U,V_2)+I(V_1;V_2|U)\\
&=I(V_1,V_2;\bby_2|U)+I(V_1;V_2|U)-I(V_2;\bby_2|U)\\
&=I(V_1,V_2;\bby_2|U)-\frac{1}{2}\log
\frac{|\bbk_2+\bbsigma_2|}{|\bbsigma_2|}\label{wei_yu_implies_1}\\
&=\frac{1}{2}\log
\frac{|\bbk_1+\bbk_2+\bbsigma_2|}{|\bbsigma_2|}-\frac{1}{2}\log
\frac{|\bbk_2+\bbsigma_2|}{|\bbsigma_2|}\\
 &=\frac{1}{2}
\log\frac{|\bbk_1+\bbk_2+\bbsigma_2|}{|\bbk_2+\bbsigma_2|}\label{ruohneg_dit_it}
\end{align}
where (\ref{wei_yu_implies_1}) comes from Theorem~1
in~\cite{Wei_Yu}. Thus, we have
\begin{align}
R_1&=I(V_1;\bby_1|U)-I(V_1;\bby_2,V_2|U)\\
&=I(\bbu_1;\bbu_1+\bbu_2+\bbn_1)-\frac{1}{2}
\log\frac{|\bbk_1+\bbk_2+\bbsigma_2|}{|\bbk_2+\bbsigma_2|}\\
&=\frac{1}{2}
\log\frac{|\bbk_1+\bbk_2+\bbsigma_1|}{|\bbk_2+\bbsigma_1|}-\frac{1}{2}
\log\frac{|\bbk_1+\bbk_2+\bbsigma_2|}{|\bbk_2+\bbsigma_2|}
\end{align}
which completes the achievability proof of $\mathcal{R}_{12}^{\rm
S-DPC-AL}(\bbs)$. Due to the symmetry, achievability of
$\mathcal{R}_{21}^{\rm S-DPC-AL}(\bbs)$ follows.

\subsection{Converse}
Since the capacity region $\mathcal{C}^{\rm AL}(\bbs)$ is convex
due to time-sharing, it can be characterized by the tangent planes
to it, i.e., by the solution of
\begin{align}
\max_{(R_0,R_1,R_2)\in\mathcal{C}^{\rm AL}(\bbs)}\mu_0 R_0+\mu_1
R_1+\mu_2 R_2
\end{align}
for $\mu_j\in[0,\infty),~j=0,1,2.$ We already have
\begin{align}
\max_{(R_0,R_1,R_2)\in \mathcal{R}^{\rm S-DPC-AL}(\bbs)}\mu_0
R_0+\mu_1 R_1+\mu_2 R_2 \leq
\max_{(R_0,R_1,R_2)\in\mathcal{C}^{\rm AL}(\bbs)}\mu_0 R_0+\mu_1
R_1+\mu_2 R_2
\end{align}
due to achievability of $\mathcal{R}_{12}^{\rm S-DPC-AL}(\bbs)$
and $\mathcal{R}_{21}^{\rm S-DPC-AL}(\bbs)$, where
$\mathcal{R}^{\rm S-DPC-AL}(\bbs)$ is given by
\begin{align}
\mathcal{R}^{\rm S-DPC-AL}(\bbs)={\rm
conv}\left(\mathcal{R}_{12}^{\rm S-DPC-AL}(\bbs)\bigcup
\mathcal{R}_{21}^{\rm S-DPC-AL}(\bbs)\right)
\end{align}
and ${\rm conv}$ is the convex hull operator. Here, we show that
\begin{align}
\max_{(R_0,R_1,R_2)\in\mathcal{C}^{\rm AL}(\bbs)}\mu_0 R_0+\mu_1
R_1+\mu_2 R_2 &\leq \max_{(R_0,R_1,R_2)\in \mathcal{R}_{12}^{\rm
S-DPC-AL}(\bbs)}\mu_0 R_0+\mu_1 R_1+\mu_2 R_2 \\
&= \max_{(R_0,R_1,R_2)\in \mathcal{R}_{21}^{\rm
S-DPC-AL}(\bbs)}\mu_0 R_0+\mu_1 R_1+\mu_2 R_2
\end{align}
to provide the converse proof. We first characterize the boundary
of $\mathcal{R}_{12}^{\rm S-DPC-AL}(\bbs)$ by studying the
following optimization problem
\begin{align}
\max_{(R_0,R_1,R_2)\in \mathcal{R}_{12}^{\rm S-DPC-AL}(\bbs)}\mu_0
R_0+\mu_1 R_1+\mu_2 R_2
\end{align}
which can be written as
\begin{align}
\max_{\substack{\bzero\preceq \bbk_j,~j=1,2
\\\bbk_1+\bbk_2\preceq \bbs}} \mu_0\min\{R_{01}^{\rm AL}(\bbk_1,\bbk_2),R_{02}^{\rm AL}(\bbk_1,\bbk_2)\}
+\mu_1 R_1^{\rm AL}(\bbk_1,\bbk_2)+\mu_2 R_2^{\rm AL}(\bbk_2)
\label{optimization}
\end{align}
Let $\bbk_1^*,\bbk_2^*$ be the maximizer of
$(\ref{optimization})$. The necessary KKT conditions that
$\bbk_1^*,\bbk_2^*$ need to satisfy are given in the following
lemma.
\begin{Lem}
\label{lemma_KKT_conditions} $\bbk_1^*,\bbk_2^*$ need to satisfy
\begin{align}
(\mu_1+\mu_2)(\bbk_1^*+\bbk_2^*+\bbsigma_1)^{-1}+\bbm_1&=(\mu_0
\lambda+\mu_2)(\bbk_1^*+\bbk_2^*+\bbsigma_1)^{-1}\nonumber\\
&\quad +(\mu_0
\bar{\lambda}+\mu_1)(\bbk_1^*+\bbk_2^*+\bbsigma_2)^{-1}+\bbm_S \label{KKT_1}\\
(\mu_1+\mu_2)(\bbk_2^*+\bbsigma_2)^{-1}+\bbm_2&=(\mu_1+\mu_2)(\bbk_2^*+\bbsigma_1)^{-1}+\bbm_1\label{KKT_2}
\end{align}
for some positive semi-definite matrices $\bbm_1,\bbm_2,\bbm_S$
such that
\begin{align}
\bbk_1^*\bbm_1&=\bbm_1\bbk_1^*=\bzero \label{KKT_3}\\
\bbk_2^*\bbm_2&=\bbm_2\bbk_2^*=\bzero\label{KKT_4} \\
(\bbs-\bbk_1^*-\bbk_2^*)\bbm_S
&=\bbm_S(\bbs-\bbk_1^*-\bbk_2^*)=\bzero\label{KKT_5}
\end{align}
and for some $\lambda=1-\bar{\lambda}$ such that it satisfies
$0\leq \lambda\leq 1$ and
\begin{align}
\lambda\left\{
\begin{array}{rcl}
&=&0\qquad {\rm if}\quad R_{01}^{\rm AL}(\bbk_1^*,\bbk_2^*)>R_{02}^{\rm AL}(\bbk_1^*,\bbk_2^*)\\
&=&1 \qquad {\rm if} \quad R_{01}^{\rm AL}(\bbk_1^*,\bbk_2^*)<R_{02}^{\rm AL}(\bbk_1^*,\bbk_2^*)\\
&\neq & 0,1 \quad  {\rm if} \quad R_{01}^{\rm
AL}(\bbk_1^*,\bbk_2^*)=R_{02}^{\rm AL}(\bbk_1^*,\bbk_2^*)
\end{array}
\right.
\end{align}
\end{Lem}
The proof of Lemma~\ref{lemma_KKT_conditions} is given in
Appendix~\ref{proof_of_lemma_KKT_conditions}.

We now use channel enhancement~\cite{Shamai_MIMO} to define a new
noise covariance matrix $\tilde{\bbsigma}$ as follows
\begin{align}
(\mu_1+\mu_2)(\bbk_2^*+\tilde{\bbsigma})^{-1}&=(\mu_1+\mu_2)(\bbk_2^*+\bbsigma_2)^{-1}+\bbm_2
\label{def_enhanced_noise}
\end{align}
This new noise covariance matrix $\tilde{\bbsigma}$ has useful
properties which are listed in the following lemma.
\begin{Lem}
\label{lemma_enhancement} We have the following facts.
\begin{itemize}
\item $\tilde{\bbsigma}\preceq \bbsigma_1,$
$\tilde{\bbsigma}\preceq \bbsigma_2$.

\item
$(\mu_1+\mu_2)(\bbk_1^*+\bbk_2^*+\tilde{\bbsigma})^{-1}=(\mu_1+\mu_2)(\bbk_1^*+\bbk_2^*+\bbsigma_1)^{-1}+\bbm_1$

\item
$(\bbk_2^*+\tilde{\bbsigma})^{-1}\tilde{\bbsigma}=(\bbk_2^*+\bbsigma_2)^{-1}\bbsigma_2$

\item
$(\bbk_1^*+\bbk_2^*+\tilde{\bbsigma})^{-1}(\bbk_2^*+\tilde{\bbsigma})=(\bbk_1^*+\bbk_2^*+\bbsigma_1)^{-1}(\bbk_2^*+\bbsigma_1)$
\end{itemize}
\end{Lem}
The proof of Lemma~\ref{lemma_enhancement} is given in
Appendix~\ref{proof_of_lemma_enhancement}. We now construct an
enhanced channel using the new covariance matrix
$\tilde{\bbsigma}$ as follows
\begin{align}
\tilde{\bby}_1&=\bbx+\tilde{\bbn} \label{enhanced_channel_1} \\
\tilde{\bby}_2&=\bbx+\tilde{\bbn} \label{enhanced_channel_2}\\
\bby_1&=\bbx+\bbn_1 \label{enhanced_channel_3}\\
\bby_2&=\bbx+\bbn_2 \label{enhanced_channel_4}
\end{align}
where $\tilde{\bbn}$ is a Gaussian random vector with a covariance
matrix $\tilde{\bbsigma}$. In the enhanced channel defined by
(\ref{enhanced_channel_1})-(\ref{enhanced_channel_4}), the
enhanced first and second users have the same observation, i.e.,
$\Pr[\tilde{\bby}_1=\tilde{\bby}_2]=1$. From now on, we denote the
observations of the enhanced first and second users by a single
random vector $\tilde{\bby}$. We now consider the following
scenario for the enhanced channel in
(\ref{enhanced_channel_1})-(\ref{enhanced_channel_4}): There are
three independent messages $(W_0,W_1,W_2)$ with rates
$(R_0,R_1,R_2)$, respectively, where the common message $W_0$ is
directed to all users, i.e., the users with observations
$\tilde{\bby}_1,\tilde{\bby}_2,\bby_1,\bby_2$; $W_1$ is the
confidential message of the enhanced first user, i.e., the one
with observation $\tilde{\bby}$, which needs to be kept perfectly
secret from the second user, i.e., the one with observation
$\bby_2$; and $W_2$ is the confidential message of the enhanced
second user, i.e., the one with observation $\tilde{\bby}$, which
needs to be kept perfectly secret from the first user, i.e., the
one with observation $\bby_1$. Here also, we measure the secrecy
of the confidential messages by normalized equivocation rates,
i.e., we require
\begin{align}
\lim_{n\rightarrow \infty} \frac{1}{n} I(W_1;\bby_2^n,W_0)=0\quad
{\rm and }\quad \lim_{n\rightarrow \infty} \frac{1}{n}
I(W_2;\bby_1^n,W_0)=0 \label{perfect_secrecy_conditions}
\end{align}
We define the capacity region of the enhanced channel in
(\ref{enhanced_channel_1})-(\ref{enhanced_channel_4}) arising from
this scenario as the convex closure of all achievable rate pairs
$(R_0,R_1,R_2)$ and denote it by $\tilde{C}(\bbs)$. Since in the
enhanced channel, the receivers to which only the common message
is sent are identical to the receivers in the original channel in
(\ref{aligned_channel_1})-(\ref{aligned_channel_2}), and the
receivers to which confidential messages are sent have better
observations with respect to the receivers in the original channel
in (\ref{aligned_channel_1})-(\ref{aligned_channel_2}), we have
$\mathcal{C}^{\rm AL}(\bbs)\subseteq \tilde{C}(\bbs)$. We next
introduce an outer bound on $\tilde{C}(\bbs)$ in the following
lemma.
\begin{Lem}
\label{lemma_outer_bound} The capacity region of the enhanced
channel in (\ref{enhanced_channel_1})-(\ref{enhanced_channel_4}),
$\tilde{C}(\bbs)$, is contained in the union of rate triples
$(R_0,R_1,R_2)$ satisfying
\begin{align}
R_0&\leq \min\{I(U;\bby_1),I(U;\bby_2)\} \\
R_1&\leq I(\bbx;\tilde{\bby}|U)-I(\bbx;\bby_2|U)\\
R_2&\leq I(\bbx;\tilde{\bby}|U)-I(\bbx;\bby_1|U)
\end{align}
for some $(U,\bbx)$ such that $U\rightarrow \bbx \rightarrow
\tilde{\bby}\rightarrow (\bby_1,\bby_2)$ and
$E\left[\bbx\bbx^\top\right]\preceq \bbs$.
\end{Lem}
The proof of this lemma is given in
Appendix~\ref{proof_of_lemma_outer_bound}. We also introduce the
following extremal inequality from~\cite{Liu_Compound}:
\begin{Lem}[\!\!\cite{Liu_Compound}, Corollary~4]
\label{lemma_extremal_liu} Let $(U,\bbx)$ be an arbitrarily
correlated random vector, where $\bbx$ has a covariance constraint
$E\left[\bbx\bbx^{\top}\right] \preceq \bbs$ and $\bbs\succ
\bzero$. Let $\tilde{\bbn},\bbn_1,\bbn_2$ be Gaussian random
vectors with covariance matrices
$\tilde{\bbsigma},\bbsigma_1,\bbsigma_2$, respectively. They are
independent of $(U,\bbx)$. Furthermore,
$\tilde{\bbsigma},\bbsigma_1,\bbsigma_2$ satisfy
$\tilde{\bbsigma}\preceq \bbsigma_j,~j=1,2$. Assume that there
exists a covariance matrix $\bbk^*$ such that $\bbk^*\preceq \bbs$
and
\begin{align}
\beta(\bbk^*+\tilde{\bbsigma})^{-1}=\sum_{j=1}^2\gamma_j
(\bbk^*+\bbsigma_j)^{-1}+\bbm_S
\end{align}
where $\beta\geq 0, \gamma_j\geq 0,~j=1,2$ and $\bbm_S$ is
positive semi-definite matrix such that
$(\bbs-\bbk^*)\bbm_S=\bzero$. Then, for any $(U,\bbx)$, we have
\begin{align}
\beta h(\bbx+\tilde{\bbn}|U)-\sum_{j=1}^2\gamma_j h(\bbx+\bbn_j|U)
\leq \frac{\beta}{2} \log |(2\pi
e)(\bbk^*+\tilde{\bbsigma})|-\sum_{j=1}^2\frac{\gamma_j}{2} \log
|(2\pi e)(\bbk^*+\bbsigma_j)|
\end{align}
\end{Lem}
We now use this lemma. For that purpose, we note that using the
second statement of Lemma~\ref{lemma_enhancement} in (\ref{KKT_1})
yields
\begin{align}
(\mu_1+\mu_2)(\bbk_1^*+\bbk_2^*+\tilde{\bbsigma})^{-1}&=(\mu_0
\lambda+\mu_2)(\bbk_1^*+\bbk_2^*+\bbsigma_1)^{-1}\nonumber\\
&\quad +(\mu_0
\bar{\lambda}+\mu_1)(\bbk_1^*+\bbk_2^*+\bbsigma_2)^{-1}+\bbm_S
\end{align}
using which in conjunction with Lemma~\ref{lemma_extremal_liu}, we
get
\begin{align}
\lefteqn{\hspace{-0.75cm}(\mu_1+\mu_2)h(\tilde{\bby}|U)-(\mu_0\lambda+\mu_2)h(\bby_1|U)-(\mu_0\bar{\lambda}+\mu_1)h(\bby_2|U)}
\nonumber \\
&\leq \frac{\mu_1+\mu_2}{2} \log |(2\pi
e)(\bbk_1^*+\bbk_2^*+\tilde{\bbsigma})|-\frac{\mu_0\lambda+\mu_2}{2}
\log|(2\pi e)(\bbk_1^*+\bbk_2^*+\bbsigma_1)| \nonumber \\
&\quad~ -\frac{\mu_0\bar{\lambda}+\mu_1}{2} \log |(2\pi
e)(\bbk_1^*+\bbk_2^*+\bbsigma_2)| \label{extremal_result}
\end{align}
which will be used subsequently.

We are now ready to complete the converse proof as follows:
\begin{align}
\lefteqn{ \max_{(R_0,R_1,R_2)\in\mathcal{C}^{\rm AL}(\bbs)}\mu_0
R_0+\mu_1 R_1+\mu_2 R_2\leq
\max_{(R_0,R_1,R_2)\in\tilde{\mathcal{C}}(\bbs)}~~\mu_0
R_0+\mu_1 R_1+\mu_2 R_2 }\label{outer_bound} \\
&\leq \max_{\substack{U\rightarrow \bbx \rightarrow
\tilde{\bby}\rightarrow \bby_1,\bby_2\\
E\left[\bbx \bbx^\top\right]\preceq \bbs}}~~\mu_0
\min\{I(U;\bby_1),I(U;\bby_2)\} +\mu_1 \left[
I(\bbx;\tilde{\bby}|U)-I(\bbx;\bby_2|U)\right]\nonumber\\
&\qquad \qquad \qquad \qquad +\mu_2 \left[I(\bbx;\tilde{\bby}|U)-I(\bbx;\bby_1|U)\right] \label{outer_bound_1} \\
&\leq \max_{\substack{U\rightarrow \bbx \rightarrow
\tilde{\bby}\rightarrow \bby_1,\bby_2\\
E\left[\bbx \bbx^\top\right]\preceq \bbs}}~~\mu_0\lambda
I(U;\bby_1)+\mu_0\bar{\lambda }I(U;\bby_2) +\mu_1 \left[
I(\bbx;\tilde{\bby}|U)-I(\bbx;\bby_2|U)\right]\nonumber\\
&\qquad \qquad \qquad \qquad +\mu_2\left[ I(\bbx;\tilde{\bby}|U)-I(\bbx;\bby_1|U) \right] \label{non_negativity}\\
&= \max_{\substack{U\rightarrow \bbx \rightarrow
\tilde{\bby}\rightarrow \bby_1,\bby_2\\
E\left[\bbx \bbx^\top\right]\preceq \bbs}}~~ \mu_0\lambda
h(\bby_1)+\mu_0\bar{\lambda
}h(\bby_2)+(\mu_1+\mu_2)h(\tilde{\bby}|U)-(\mu_0\lambda+\mu_2)h(\bby_1|U)\nonumber\\
&\qquad \qquad \qquad \qquad
-(\mu_0\bar{\lambda}+\mu_1)h(\bby_2|U) -\frac{\mu_1}{2}
\log\frac{|\tilde{\bbsigma}|}{|\bbsigma_2|}-\frac{\mu_2}{2}
\log\frac{|\tilde{\bbsigma}|}{|\bbsigma_1|} \\
&\leq \frac{\mu_0\lambda}{2} \log |(2\pi e)(\bbs+\bbsigma_1)|
+\frac{\mu_0\bar{\lambda }}{2}  \log|(2\pi
e)(\bbs+\bbsigma_2)|\nonumber\\
&\quad + \max_{\substack{U\rightarrow \bbx \rightarrow
\tilde{\bby}\rightarrow \bby_1,\bby_2\\
E\left[\bbx \bbx^\top\right]\preceq \bbs}}~~
(\mu_1+\mu_2)h(\tilde{\bby}|U)-(\mu_0\lambda+\mu_2)h(\bby_1|U)
-(\mu_0\bar{\lambda}+\mu_1)h(\bby_2|U) \nonumber\\
&\quad  -\frac{\mu_1}{2}
\log\frac{|\tilde{\bbsigma}|}{|\bbsigma_2|}-\frac{\mu_2}{2}
\log\frac{|\tilde{\bbsigma}|}{|\bbsigma_1|} \label{max_entropy_theorem} \\
&\leq \frac{\mu_0\lambda}{2} \log |(2\pi e)(\bbs+\bbsigma_1)|
+\frac{\mu_0\bar{\lambda }}{2}  \log|(2\pi
e)(\bbs+\bbsigma_2)|\nonumber\\
&\quad + \frac{(\mu_1+\mu_2)}{2} \log |(2\pi
e)(\bbk_1^*+\bbk_2^*+\tilde{\bbsigma})|-\frac{(\mu_0\lambda+\mu_2)}{2}
\log |(2\pi e)(\bbk_1^*+\bbk_2^*+\bbsigma_1)| \nonumber \\
&\quad -\frac{(\mu_0\bar{\lambda}+\mu_1)}{2}\log|(2\pi
e)(\bbk_1^*+\bbk_2^*+\bbsigma_2)|  -\frac{\mu_1}{2}
\log\frac{|\tilde{\bbsigma}|}{|\bbsigma_2|}-\frac{\mu_2}{2}
\log\frac{|\tilde{\bbsigma}|}{|\bbsigma_1|}
\label{extremal_result_implies}\\
&= \frac{\mu_0\lambda}{2} \log
\frac{|\bbs+\bbsigma_1|}{|\bbk_1^*+\bbk_2^*+\bbsigma_1|}
+\frac{\mu_0\bar{\lambda }}{2} \log\frac{|\bbs+\bbsigma_2|}{|\bbk_1^*+\bbk_2^*+\bbsigma_2|}\nonumber\\
&\quad + \frac{\mu_1}{2} \log
\frac{|(\bbk_1^*+\bbk_2^*+\tilde{\bbsigma})\bbsigma_2|}{|(\bbk_1^*+\bbk_2^*+\bbsigma_2)\tilde{\bbsigma}|}
+ \frac{\mu_2}{2} \log
\frac{|(\bbk_1^*+\bbk_2^*+\tilde{\bbsigma})\bbsigma_1|}{|(\bbk_1^*+\bbk_2^*+\bbsigma_1)\tilde{\bbsigma}|}\\
&=\mu_0 \min\{R_{01}^{\rm AL}(\bbk_1^*,\bbk_2^*),R_{02}^{\rm
AL}(\bbk_1^*,\bbk_2^*)\} + \frac{\mu_1}{2} \log
\frac{|(\bbk_1^*+\bbk_2^*+\tilde{\bbsigma})\bbsigma_2|}{|(\bbk_1^*+\bbk_2^*+\bbsigma_2)\tilde{\bbsigma}|}
\nonumber\\
&\quad + \frac{\mu_2}{2} \log
\frac{|(\bbk_1^*+\bbk_2^*+\tilde{\bbsigma})\bbsigma_1|}{|(\bbk_1^*+\bbk_2^*+\bbsigma_1)\tilde{\bbsigma}|}\label{back_to_common_messages}\\
&=\mu_0 \min\{R_{01}^{\rm AL}(\bbk_1^*,\bbk_2^*),R_{02}^{\rm
AL}(\bbk_1^*,\bbk_2^*)\} +\mu_1 R_1^{\rm
AL}(\bbk_1^*,\bbk_2^*)+\mu_2 R_2^{\rm
AL}(\bbk_2^*)\label{lemma_enhancement_implies_5}
\end{align}
where (\ref{outer_bound}) comes from the fact that
$\mathcal{C}^{\rm AL}(\bbs)\subseteq \tilde{\mathcal{C}}(\bbs)$,
(\ref{outer_bound_1}) is due to Lemma~\ref{lemma_outer_bound},
(\ref{non_negativity}) results from the fact that $0\leq
\lambda=1-\bar{\lambda} \leq 1$, (\ref{max_entropy_theorem}) is
due to the maximum entropy theorem,
(\ref{extremal_result_implies}) comes from
(\ref{extremal_result}), (\ref{back_to_common_messages}) results
from
\begin{align}
\lambda R_{01}^{\rm AL}(\bbk_1^*,\bbk_2^*)+\bar{\lambda}
R_{02}^{\rm AL}(\bbk_1^*,\bbk_2^*)=\min\{R_{01}^{\rm
AL}(\bbk_1^*,\bbk_2^*),R_{02}^{\rm AL}(\bbk_1^*,\bbk_2^*)\}
\end{align}
and (\ref{lemma_enhancement_implies_5}) will be shown next. We
first note the following
\begin{align}
R_1^{\rm AL}(\bbk_1^*,\bbk_2^*)&=\frac{1}{2} \log
\frac{|\bbk_1^*+\bbk_2^*+\bbsigma_1|}{|\bbk_2^*+\bbsigma_1|}
-\frac{1}{2}\log
\frac{|\bbk_1^*+\bbk_2^*+\bbsigma_2|}{|\bbk_2^*+\bbsigma_2|} \\
&=\frac{1}{2}\log
\frac{|\bbk_1^*+\bbk_2^*+\tilde{\bbsigma}|}{|\bbk_2^*+\tilde{\bbsigma}|}
-\frac{1}{2}\log
\frac{|\bbk_1^*+\bbk_2^*+\bbsigma_2|}{|\bbk_2^*+\bbsigma_2|}\label{lemma_enhancement_implies_1}\\
&= \frac{1}{2}
 \log
\frac{|(\bbk_1^*+\bbk_2^*+\tilde{\bbsigma})\bbsigma_2|}{|(\bbk_1^*+\bbk_2^*+\bbsigma_2)\tilde{\bbsigma}|}
\label{lemma_enhancement_implies_2}
\end{align}
where (\ref{lemma_enhancement_implies_1}) is due to the fourth
statement of Lemma~\ref{lemma_enhancement} and
(\ref{lemma_enhancement_implies_2}) comes from the third statement
of Lemma~\ref{lemma_enhancement}. We next note the following
identity
\begin{align}
R_2^{\rm AL}(\bbk_2^*)&=\frac{1}{2}\log
\frac{|\bbk_2^*+\bbsigma_2|}{|\bbsigma_2|}-\frac{1}{2}\log
\frac{|\bbk_2^*+\bbsigma_1|}{|\bbsigma_1|}\\
&=\frac{1}{2}\log
\frac{|\bbk_2^*+\tilde{\bbsigma}|}{|\tilde{\bbsigma}|}-\frac{1}{2}\log
\frac{|\bbk_2^*+\bbsigma_1|}{|\bbsigma_1|}
\label{lemma_enhancement_implies_3}\\
&= \log
\frac{|(\bbk_1^*+\bbk_2^*+\tilde{\bbsigma})\bbsigma_1|}{|(\bbk_1^*+\bbk_2^*+\bbsigma_1)\tilde{\bbsigma}|}
\label{lemma_enhancement_implies_4}
\end{align}
where (\ref{lemma_enhancement_implies_3}) is due to the third
statement of Lemma~\ref{lemma_enhancement}, and
(\ref{lemma_enhancement_implies_4}) comes from the fourth
statement of Lemma~\ref{lemma_enhancement}. Identities in
(\ref{lemma_enhancement_implies_2}) and
(\ref{lemma_enhancement_implies_4}) give
(\ref{lemma_enhancement_implies_5}).

Thus, in the view of (\ref{lemma_enhancement_implies_5}), we have
shown that
\begin{align}
\max_{(R_0,R_1,R_2)\in\mathcal{C}^{\rm AL}(\bbs)}\mu_0
R_0+\mu_1R_1+\mu_2 R_2 &=
\max_{(R_0,R_1,R_2)\in\mathcal{R}_{12}^{\rm S-DPC-AL}(\bbs)} \mu_0
R_0+\mu_1 R_1+\mu_2 R_2
\end{align}
Similarly, we can show the following
\begin{align}
\max_{(R_0,R_1,R_2)\in\mathcal{C}^{\rm AL}(\bbs)}\mu_0
R_0+\mu_1R_1+\mu_2 R_2 &=
\max_{(R_0,R_1,R_2)\in\mathcal{R}_{21}^{\rm S-DPC-AL}(\bbs)} \mu_0
R_0+\mu_1 R_1+\mu_2 R_2
\end{align}
completing the converse proof.

\section{Proof of Theorem~\ref{theorem_main_result} for the General Case}

We now prove Theorem~\ref{theorem_main_result} for the general
channel model in
(\ref{general_gaussian_mimo_1})-(\ref{general_gaussian_mimo_2}).
Achievability of Theorem~\ref{theorem_main_result} for the general
channel model in
(\ref{general_gaussian_mimo_1})-(\ref{general_gaussian_mimo_2})
can be shown as we did for the aligned case in the previous
section. In particular, the only difference of the achievability
proof for the general channel model in
(\ref{general_gaussian_mimo_1})-(\ref{general_gaussian_mimo_2})
from the achievability proof for the aligned case will be the
selection of the precoding matrix $\bba$, which needs to be chosen
as $\bba=\bbk_2\bbh_2^\top
(\bbsigma_2+\bbh_2\bbk_2\bbh_2^\top)^{-1}\bbh_2$ in this general
case. Thus, in the rest of this section, we consider the converse
proof. For that purpose, we follow the analysis in Section~V.B
of~\cite{Shamai_MIMO} and Section~7.1 of~\cite{MIMO_BC_Secrecy} in
conjunction with the capacity result obtained for the aligned case
in the previous section. To this end, we first note that,
following the approaches in Section~V.B of~\cite{Shamai_MIMO} and
Section~7.1 of~\cite{MIMO_BC_Secrecy}, it can be shown that a new
channel can be constructed from any channel described by
(\ref{general_gaussian_mimo_1})-(\ref{general_gaussian_mimo_2}),
such that the new channel has the same capacity region as the
original one, and in the new channel, both receivers have the same
number of antennas as the transmitter, i.e., $r_1=r_2=t$. Thus,
without loss of generality, we assume that $r_1=r_2=t$. We next
apply singular-value decomposition to the channel gain matrices
$\bbh_1,\bbh_2$ as follows
\begin{align}
\bbh_j=\bbu_j\bblambda_j\bbv_j^\top,\quad j=1,2
\end{align}
where $\bbu_j,\bbv_j$ are $t\times t$ orthogonal matrices, and
$\bblambda_j$ is a diagonal matrix. We now define a new Gaussian
MIMO broadcast channel as follows
\begin{align}
\overline{\bby}_1&=\overline{\bbh}_1\bbx+\bbn_1 \label{channel_alpha_1}\\
\overline{\bby}_2&=\overline{\bbh}_2\bbx+\bbn_2
\label{channel_alpha_2}
\end{align}
where $\overline{\bbh}_j$ is defined as
\begin{align}
\overline{\bbh}_j=\bbu_j (\bblambda_j+\alpha\bbi)\bbv_j^\top
\end{align}
for some $\alpha>0$. We denote the capacity region of the channel
defined in (\ref{channel_alpha_1})-(\ref{channel_alpha_2}) by
$\mathcal{C}_{\alpha}(\bbs)$, and achievable rate regions for this
channel by $\mathcal{R}_{12,\alpha}^{\rm
S-DPC}(\bbs),\mathcal{R}_{21,\alpha}^{\rm S-DPC}(\bbs)$. Since
$\overline{\bbh}_1,\overline{\bbh}_2$ are invertible, the capacity
region of the channel in
(\ref{channel_alpha_1})-(\ref{channel_alpha_2}) is equal to the
capacity region of the following aligned channel
\begin{align}
\overline{\overline{\bby}}_1&=\bbx+\overline{\bbh}_1^{~-1}\bbn_1 \label{channel_alpha_again_1}\\
\overline{\overline{\bby}}_2&=\bbx+\overline{\bbh}_2 ^{~-1} \bbn_2
\label{channel_alpha_again_2}
\end{align}
Thus, using the capacity result for the aligned case, which was
proved in the previous section, we get
\begin{align}
\mathcal{C}_{\alpha}(\bbs)=\mathcal{R}_{12,\alpha}^{\rm
S-DPC}(\bbs)=\mathcal{R}_{21,\alpha}^{\rm S-DPC}(\bbs)
\end{align}
We next show the following inclusion
\begin{align}
\mathcal{C}(\bbs) \subseteq \lim_{\alpha\rightarrow
0}\mathcal{C}_{\alpha}(\bbs) \label{inclusion}
\end{align}
To this end, assume that $(R_0,R_1,R_2)$ is achievable in the
channel given by
(\ref{general_gaussian_mimo_1})-(\ref{general_gaussian_mimo_2}),
i.e., $(R_0,R_1,R_2)\in\mathcal{C}(\bbs)$. To prove the inclusion
in (\ref{inclusion}), we need to show that
$(R_0,R_1,R_2)\in\lim_{\alpha\rightarrow
0}\mathcal{C}_{\alpha}(\bbs)$. To this end, we note the following
Markov chains
\begin{align}
\bbx\rightarrow \overline{\bby}_j \rightarrow \bby_j,\quad j=1,2
\label{auxiliary_mc_1}
\end{align}
which imply that if the message triple $(W_0,W_1,W_2)$ with rates
$(R_0,R_1,R_2)$ is transmitted with a vanishingly small
probability of error in the original channel given by
(\ref{general_gaussian_mimo_1})-(\ref{general_gaussian_mimo_2}),
they will be transmitted with a vanishingly small probability of
error in the channel given by
(\ref{channel_alpha_1})-(\ref{channel_alpha_2}) as well. In other
words, each receiver in the channel given by
(\ref{channel_alpha_1})-(\ref{channel_alpha_2}) will decode the
messages intended to itself. However, we still need to check the
secrecy requirements on the confidential messages $W_1,W_2$. We
first check the secrecy of the first user's confidential message
as follows
\begin{align}
\lim_{n\rightarrow
\infty}\frac{1}{n}I(W_1;\overline{\bby}_2^n,W_0,W_2)&=\lim_{n\rightarrow
\infty}
\frac{1}{n}I(W_1;\overline{\bby}_2^n,W_0,W_2)-\frac{1}{n}I(W_1;\bby_2^n,W_0,W_2)
\label{check_secrecy_1}
\end{align}
where we used the fact that since
$(R_0,R_1,R_2)\in\mathcal{C}(\bbs)$, we have
\begin{align}
\lim_{n\rightarrow \infty} \frac{1}{n}I(W_1;\bby_2^n,W_0,W_2)=0
\end{align}
We now bound the term on the right hand-side of as follows
(\ref{check_secrecy_1})
\begin{align}
\lefteqn{\hspace{-1cm}I(W_1;\overline{\bby}_2^n,W_0,W_2)-I(W_1;\bby_2^n,W_0,W_2)}\nonumber\\
&=I(W_1;\overline{\bby}_2^n|W_0,W_2)-I(W_1;\bby_2^n|W_0,W_2)\\
&= I(W_1;\overline{\bby}_2^n|W_0,W_2,\bby_2^n)\label{degradedness_imply_again}\\
&= \sum_{i=1}^n
I(W_1;\overline{\bby}_{2,i}|W_0,W_2,\bby_2^n,\overline{\bby}_{2}^{i-1})\\
&\leq  \sum_{i=1}^n h(\overline{\bby}_{2,i}|\bby_{2,i})-
h(\overline{\bby}_{2,i}|W_0,W_2,\bby_2^n,\overline{\bby}_{2}^{i-1},W_1,\bbx_i)
\label{conditioning_cannot_1} \\
&=  \sum_{i=1}^n
I(\bbx_i;\overline{\bby}_{2,i}|\bby_{2,i})\label{memoryless_2}  \\
&=  \sum_{i=1}^n
I(\bbx_i;\overline{\bby}_{2,i})-I(\bbx_i;\bby_{2,i})\label{degradedness_cond_imply_x}\\
&= \sum_{i=1}^n h(\overline{\bby}_{2,i})-h(\bby_{2,i}) \\
&\leq \sum_{i=1}^n \frac{1}{2}\log \frac{\left|\overline{\bbh}_2
E\left[\bbx_i\bbx_i^\top\right]\overline{\bbh}_2^\top+\bbsigma_2\right|}{\left|\bbh_2
E\left[\bbx_i\bbx_i^\top\right]\bbh_2^\top+\bbsigma_2\right|}
\label{worst_additive_lemma}\\
&\leq  \frac{n}{2}\log \frac{\left|\overline{\bbh}_2
\left(\sum_{i=1}^n
\frac{1}{n}E\left[\bbx_i\bbx_i^\top\right]\right)\overline{\bbh}_2^\top+\bbsigma_2\right|}{\left|\bbh_2
\left(\sum_{i=1}^n
\frac{1}{n}E\left[\bbx_i\bbx_i^\top\right]\right)\bbh_2^\top+\bbsigma_2\right|}
\label{concavity} \\
&\leq \frac{n}{2}\log \frac{\left|\overline{\bbh}_2 \bbs
\overline{\bbh}_2^\top+\bbsigma_2\right|}{\left|\bbh_2 \bbs
\bbh_2^\top+\bbsigma_2\right|} \label{monotonicity}
\end{align}
where (\ref{degradedness_imply_again}) is due to the Markov chain
in (\ref{auxiliary_mc_1}), (\ref{conditioning_cannot_1}) comes
from the fact that conditioning cannot increase entropy,
(\ref{memoryless_2}) is due to the fact that the channel is
memoryless, (\ref{degradedness_cond_imply_x}) results from the
Markov chain in (\ref{auxiliary_mc_1}), and
(\ref{worst_additive_lemma}) can be shown by using the worst
additive noise lemma in~\cite{Diggavi_Cover,Ihara}. Before showing
the steps in (\ref{concavity}) and (\ref{monotonicity}), we note
that the following function
\begin{align}
\log \frac{\left|\overline{\bbh}_2 \bbk
\overline{\bbh}_2^\top+\bbsigma_2\right|}{\left|\bbh_2 \bbk
\bbh_2^\top+\bbsigma_2\right|} \label{dummy_function}
\end{align}
is concave and monotonically increasing in positive semi-definite
matrices $\bbk$, see Lemma~4 in~\cite{Yingbin_Compound}. Thus,
(\ref{concavity}) follows from the Jensen's inequality by noting
the concavity of the function in (\ref{dummy_function}) and
(\ref{monotonicity}) comes from the monotonicity of the function
in (\ref{dummy_function}) and the covariance constraint on the
channel input. Hence, using (\ref{monotonicity}) in
(\ref{check_secrecy_1}), we have
\begin{align}
\lim_{n\rightarrow
\infty}\frac{1}{n}I(W_1;\overline{\bby}_2^n,W_0,W_2)&\leq
\frac{1}{2}\log \frac{\left|\overline{\bbh}_2 \bbs
\overline{\bbh}_2^\top+\bbsigma_2\right|}{\left|\bbh_2 \bbs
\bbh_2^\top+\bbsigma_2\right|}
\end{align}
where the right hand-side vanishes as $\alpha\rightarrow 0$, i.e.,
\begin{align}
\lim_{\alpha\rightarrow 0}\frac{1}{2}\log
\frac{\left|\overline{\bbh}_2 \bbs
\overline{\bbh}_2^\top+\bbsigma_2\right|}{\left|\bbh_2 \bbs
\bbh_2^\top+\bbsigma_2\right|} =0
\end{align}
due to the continuity of $\log|\cdot|$ in positive semi-definite
matrices and $\lim_{\alpha\rightarrow 0}\overline{\bbh}_2=\bbh_2$.
Thus, we have shown that if a confidential message $W_1$ with rate
$R_1$ can be transmitted in perfect secrecy in the original
channel given by
(\ref{general_gaussian_mimo_1})-(\ref{general_gaussian_mimo_2}),
we have
\begin{align}
\lim_{\alpha \rightarrow 0}\lim_{n\rightarrow
\infty}\frac{1}{n}I(W_1;\overline{\bby}_2^n,W_0,W_2)&=0
\label{check_secrecy_2}
\end{align}
Similarly, if a confidential message $W_2$ with rate $R_2$ can be
transmitted in perfect secrecy in the original channel given by
(\ref{general_gaussian_mimo_1})-(\ref{general_gaussian_mimo_2}),
we have
\begin{align}
\lim_{\alpha \rightarrow 0}\lim_{n\rightarrow
\infty}\frac{1}{n}I(W_2;\overline{\bby}_1^n,W_0,W_1)&=0
\label{check_secrecy_3}
\end{align}
These two conditions in (\ref{check_secrecy_2}) and
(\ref{check_secrecy_3}) enable us to conclude that if
$(R_0,R_1,R_2)\in\mathcal{C}(\bbs)$, we also have
$(R_0,R_1,R_2)\in\lim_{\alpha\rightarrow
0}\mathcal{C}_{\alpha}(\bbs)$. Thus, we have shown that
\begin{align}
\mathcal{C}(\bbs)\subseteq \lim_{\alpha\rightarrow
0}\mathcal{C}_{\alpha}(\bbs)=\lim_{\alpha\rightarrow
0}\mathcal{R}_{12,\alpha}^{\rm
S-DPC}(\bbs)=\lim_{\alpha\rightarrow
0}\mathcal{R}_{21,\alpha}^{\rm S-DPC}(\bbs) \label{bound}
\end{align}
where we have
\begin{align}
\lim_{\alpha\rightarrow 0}\mathcal{R}_{12,\alpha}^{\rm
S-DPC}(\bbs)&=\mathcal{R}_{12}^{\rm
S-DPC}(\bbs)\label{convergence_1}\\
\lim_{\alpha\rightarrow 0}\mathcal{R}_{21,\alpha}^{\rm
S-DPC}(\bbs)&=\mathcal{R}_{21}^{\rm S-DPC}(\bbs)
\label{convergence_2}
\end{align}
due to the continuity of the rate expressions in
$\mathcal{R}_{12,\alpha}^{\rm S-DPC}(\bbs)$ and
$\mathcal{R}_{21,\alpha}^{\rm S-DPC}(\bbs)$ in $\alpha$. Since
$\mathcal{R}_{12}^{\rm S-DPC}(\bbs)$ and $\mathcal{R}_{21}^{\rm
S-DPC}(\bbs)$ are achievable in the channel defined by
(\ref{general_gaussian_mimo_1})-(\ref{general_gaussian_mimo_2}),
we have
\begin{align}
\mathcal{C}(\bbs)=\mathcal{R}_{12}^{\rm
S-DPC}(\bbs)=\mathcal{R}_{21}^{\rm S-DPC}(\bbs)
\end{align}
in the view of (\ref{bound})-(\ref{convergence_2}); completing the
proof.

\section{Connections to the Gaussian MIMO Broadcast Channel with Common and Private Messages}
Here, we provide intuitive explanations for the two facts that
Theorem~\ref{theorem_main_result} reveals: i) The achievable rate
region does not depend on the encoding order used in S-DPC, i.e.,
$\mathcal{R}_{12}^{\rm S-DPC}(\bbs)\break =\mathcal{R}_{21}^{\rm
S-DPC}(\bbs)$; and ii) the capacity region of the Gaussian MIMO
broadcast channel with common and confidential messages can be
completely characterized, although the capacity region of the its
non-confidential counterpart, i.e., the Gaussian MIMO broadcast
channel with common and private messages, is not known completely.

In the Gaussian MIMO broadcast channel with common and private
messages, there are again three messages $W_0,W_1,W_2$ with rates
$R_0,R_1,R_2$, respectively, such that $W_0$ is again sent to both
users, $W_1$ (resp. $W_2$) is again directed to only the first
(resp. second) user, however, there are no secrecy constraints on
$W_1,W_2$. The capacity region of the Gaussian MIMO broadcast
channel with common and private messages will be denoted by
$\mathcal{C}^{\rm NS}(\bbs)$. The achievable rate region for the
Gaussian MIMO broadcast channel with common and private messages
that can be obtained by using DPC will be denoted by
$\mathcal{R}_{12}^{\rm NS-DPC}(\bbs),\mathcal{R}_{21}^{\rm
NS-DPC}(\bbs)$ (depending on the encoding order), where
$\mathcal{R}_{12}^{\rm NS-DPC}(\bbs)$ is given by the rate triples
$(R_0,R_1,R_2)$ satisfying
\begin{align}
R_0&\leq \min\{R_{01}^{\rm NS}(\bbk_1,\bbk_2),R_{02}^{\rm
NS}(\bbk_1,\bbk_2)\} \\
R_1&\leq R_{1}^{\rm NS}(\bbk_1,\bbk_2) \\
R_2&\leq R_{2}^{\rm NS}(\bbk_2)
\end{align}
for some positive semi-definite matrices $\bbk_1,\bbk_2$ such that
$\bbk_1+\bbk_2\preceq \bbs$, and $ \{R_{0j}^{\rm
NS}(\bbk_1,\bbk_2)\}_{j=1}^2,\break R_{1}^{\rm
NS}(\bbk_1,\bbk_2),R_{2}^{\rm NS}(\bbk_1,\bbk_2)$ are defined as
\begin{align}
R_{0j}^{\rm NS}(\bbk_1,\bbk_2)&=\frac{1}{2}
\log\frac{|\bbs+\bbsigma_j|}{|\bbk_1+\bbk_2+\bbsigma_j|},\quad
j=1,2 \\
R_{1}^{\rm NS}(\bbk_1,\bbk_2)&=\frac{1}{2} \log
\frac{|\bbk_1+\bbk_2+\bbsigma_1|}{|\bbk_2+\bbsigma_1|} \\
R_{2}^{\rm NS}(\bbk_1,\bbk_2)&=\frac{1}{2} \log
\frac{|\bbk_2+\bbsigma_2|}{|\bbsigma_2|}
\end{align}
Moreover, $\mathcal{R}_{21}^{\rm NS-DPC}(\bbs)$ can be obtained
from $\mathcal{R}_{12}^{\rm NS-DPC}(\bbs)$ by swapping the
subscripts 2 and 1. We now state a result of~\cite{hannan_thesis}
on the capacity region of the Gaussian MIMO broadcast channel with
common and private messages: For a given common message rate
$R_0$, the private message sum rate capacity, i.e., $R_1+R_2$, is
achieved by both $\mathcal{R}_{12}^{\rm NS}(\bbs)$ and
$\mathcal{R}_{21}^{\rm NS}(\bbs)$. This result can also be stated
as follows
\begin{align}
\max_{(R_0,R_1,R_2)\in\mathcal{C}^{\rm NS}(\bbs)} \mu_0^\prime
R_0+ \mu_1^{\prime} R_1 +\mu_2^{\prime} R_2 &=
\max_{(R_0,R_1,R_2)\in \mathcal{R}_{12}^{\rm NS-DPC}(\bbs)}
\mu_0^\prime R_0+ \mu_1^{\prime} R_1 +\mu_2^{\prime} R_2 \label{hannan_did_it_1}\\
&= \max_{(R_0,R_1,R_2)\in \mathcal{R}_{21}^{\rm NS-DPC}(\bbs)}
\mu_0^\prime R_0+ \mu_1^{\prime} R_1 +\mu_2^{\prime} R_2
\label{hannan_did_it_2}
\end{align}
for $\mu_1^\prime=\mu_2^\prime=\mu^\prime$.  This result is
crucial to understand the aforementioned two points suggested by
Theorem~\ref{theorem_main_result}, which will be explained next
using (\ref{hannan_did_it_1})-(\ref{hannan_did_it_2}).

In the proof of Theorem~\ref{theorem_main_result}, first, we
characterize the boundary of $\mathcal{R}_{12}^{\rm S-DPC}(\bbs)$
by finding the properties of the covariance matrices that achieve
the boundary of $\mathcal{R}_{12}^{\rm S-DPC}(\bbs)$, see
Lemma~\ref{lemma_KKT_conditions}. According to
Lemma~\ref{lemma_KKT_conditions}, the boundary of
$\mathcal{R}_{12}^{\rm S-DPC}(\bbs)$ can be achieved by using the
covariance matrices $\bbk_1^*,\bbk_2^*$ satisfying
\begin{align}
(\mu_1+\mu_2)(\bbk_1^*+\bbk_2^*+\bbsigma_1)^{-1}+\bbm_1&=(\mu_0
\lambda+\mu_2)(\bbk_1^*+\bbk_2^*+\bbsigma_1)^{-1}\nonumber\\
&\quad +(\mu_0
\bar{\lambda}+\mu_1)(\bbk_1^*+\bbk_2^*+\bbsigma_2)^{-1}+\bbm_S \label{KKT_again_1}\\
(\mu_1+\mu_2)(\bbk_2^*+\bbsigma_2)^{-1}+\bbm_2&=(\mu_1+\mu_2)(\bbk_2^*+\bbsigma_1)^{-1}+\bbm_1
\label{KKT_again_2}
\end{align}
On the other hand, using these covariance matrices, we can also
achieve the boundary points of $\mathcal{R}_{12}^{\rm
NS-DPC}(\bbs)$, which are actually on the boundary of the capacity
region $\mathcal{C}^{\rm NS}(\bbs)$ as well, and are the private
message sum rate capacity points for a given common message rate.
To see this point, we define $\mu^\prime=\mu_1+\mu_2,\mu_0^\prime
=\mu_0+\mu_1+\mu_2$ and $\gamma=\frac{\mu_0
\lambda+\mu_2}{\mu_0+\mu_1+\mu_2}$, i.e.,
$\bar{\gamma}=1-\gamma=\frac{\mu_0
\bar{\lambda}+\mu_1}{\mu_0+\mu_1+\mu_2}$. Thus, the conditions in
(\ref{KKT_again_1})-(\ref{KKT_again_2}) can be written as
\begin{align}
\mu^\prime (\bbk_1^*+\bbk_2^*+\bbsigma_1)^{-1}+\bbm_1&=\mu_0^\prime\gamma(\bbk_1^*+\bbk_2^*+\bbsigma_1)^{-1} +\mu_0^\prime \bar{\gamma}(\bbk_1^*+\bbk_2^*+\bbsigma_2)^{-1}+\bbm_S \label{KKT_again_1_1}\\
\mu^\prime(\bbk_2^*+\bbsigma_2)^{-1}+\bbm_2&=\mu^\prime
(\bbk_2^*+\bbsigma_1)^{-1}+\bbm_1 \label{KKT_again_2_1}
\end{align}
which are the necessary conditions that the following problem
needs to satisfy
\begin{align}
\max_{(R_0,R_1,R_2)\in \mathcal{R}_{12}^{\rm NS-DPC}(\bbs)}
\mu_0^\prime R_0+ \mu^{\prime} (R_1 +R_2) \label{intuition}
\end{align}
On the other hand, due to
(\ref{hannan_did_it_1})-(\ref{hannan_did_it_2}), we know that the
solution of (\ref{intuition}) gives us the private message sum
rate capacity for a given common message rate, i.e., the points
that achieve the maximum in (\ref{intuition}) are on the boundary
of the capacity region $\mathcal{C}^{\rm NS}(\bbs)$. Furthermore,
the maximum value in (\ref{intuition}) can also be achieved by
using the other possible encoding order, i.e.,
\begin{align}
\max_{(R_0,R_1,R_2)\in \mathcal{R}_{12}^{\rm NS-DPC}(\bbs)}
\mu_0^\prime R_0+ \mu^{\prime} (R_1 +R_2)= \max_{(R_0,R_1,R_2)\in
\mathcal{R}_{21}^{\rm NS-DPC}(\bbs)} \mu_0^\prime R_0+
\mu^{\prime} (R_1 +R_2)
\end{align}
Thus, this discussion reveals that there is a one-to-one
correspondence between any rate triple on the boundary of
$\mathcal{R}_{12}^{\rm S-DPC}(\bbs)$ and the private message sum
rate capacity points on $\mathcal{C}^{\rm NS}(\bbs)$. Hence, the
boundary of $\mathcal{R}_{12}^{\rm S-DPC}(\bbs)$, similarly
$\mathcal{R}_{21}^{\rm S-DPC}(\bbs)$, can be constructed by
considering the private message sum rate capacity points on
$\mathcal{C}^{\rm NS}(\bbs)$. This connection between the private
message sum rate capacity points and the boundaries of
$\mathcal{R}_{12}^{\rm S-DPC}(\bbs)$, $\mathcal{R}_{21}^{\rm
S-DPC}(\bbs)$ intuitively explains the two facts suggested by
Theorem~\ref{theorem_main_result}: i) The achievable rate region
for the Gaussian MIMO broadcast channel with common and
confidential messages is invariant with respect to the encoding
order, i.e., $\mathcal{R}_{12}^{\rm
S-DPC}(\bbs)=\mathcal{R}_{21}^{\rm S-DPC}(\bbs)$ because the
boundaries of these two regions correspond to those points on the
DPC region for the Gaussian MIMO broadcast channel with common and
private messages, for which encoding order does not matter either;
and ii) we can obtain the entire capacity region of the Gaussian
MIMO broadcast channel with common and confidential messages,
although the capacity region of its non-confidential counterpart
is not known completely. The reason is that the boundary of the
capacity region of the Gaussian MIMO broadcast channel with common
and confidential messages comes from those points on the boundary
of the DPC region of its non-confidential counterpart, which are
known to be tight, i.e., which are known to be on the boundary of
the capacity region of the Gaussian MIMO broadcast channel with
common and private messages.

\section{Conclusions}
We study the Gaussian MIMO broadcast channel with common and
confidential messages, and obtain the entire capacity region. We
show that a variant of the S-DPC scheme proposed
in~\cite{Ruoheng_MIMO_BC} is capacity-achieving. We provide the
converse proof by using channel enhancement~\cite{Shamai_MIMO} and
an extremal inequality from~\cite{Liu_Compound}. We also uncover
the connections between the Gaussian MIMO broadcast channel with
common and confidential messages and its non-confidential
counterpart, i.e., the Gaussian MIMO broadcast channel with common
and private messages, to provide further insight into capacity
result we obtained.

\appendices
\appendixpage

\section{Proof of Lemma~\ref{lemma_KKT_conditions}}
\label{proof_of_lemma_KKT_conditions}

Since the program in (\ref{optimization}) is not necessarily
convex, the KKT conditions are necessary but not sufficient. We
first rewrite the program in (\ref{optimization}) as follows
\begin{align}
&\max_{\substack{\bzero\preceq \bbk_j,~j=1,2 \\
\bbk_1+\bbk_2\preceq \bbs\\a }}~~\mu_0 a+\mu_1 R_1^{\rm
AL}(\bbk_1,\bbk_2)+\mu_2 R_2^{\rm AL}(\bbk_2) \nonumber \\
&\quad {\rm s.t.}\qquad \quad  R_{01}^{\rm AL}(\bbk_1,\bbk_2)\geq a \nonumber \\
&\qquad \qquad \quad \hspace{0.1cm} R_{02}^{\rm
AL}(\bbk_1,\bbk_2)\geq a \label{optimization_rewritten}
\end{align}
where we introduce an additional variable $a$. Thus, the
optimization in (\ref{optimization_rewritten}) is over three
variables $a,\bbk_1,\bbk_2$. The Lagrangian of
(\ref{optimization_rewritten}) is given by
\begin{align}
\mathcal{L}&= \mu_0 a+\mu_1 R_1^{\rm AL}(\bbk_1,\bbk_2)+\mu_2
R_2^{\rm AL}(\bbk_2)+ \mu_0\sum_{j=1}^2\lambda_j (R_{0j}^{\rm
AL}(\bbk_1,\bbk_2)-a) + {\rm tr}(\bbk_1\bbm_1)\nonumber\\
&\quad +{\rm tr}(\bbk_2\bbm_2)+{\rm
tr}((\bbs-\bbk_1-\bbk_2)\bbm_S)
\end{align}
where $\bbm_1,\bbm_2,\bbm_S$ are positive semi-definite matrices
and $\lambda_j\geq 0,~j=1,2$. Let $(a^*,\bbk_1^*,\bbk_2^*)$ be the
maximizer for (\ref{optimization_rewritten}). The necessary KKT
conditions that they need to satisfy are given as follows
\begin{align}
\frac{\partial{\mathcal{L}}}{\partial a} \mid_{a=a^*}&=0 \label{proof_KKT_1}\\
\nabla_{\bbk_1}{\mathcal{L}}\mid_{\bbk_1=\bbk_1^*}&=\bzero \label{proof_KKT_2}\\
\nabla_{\bbk_2}{\mathcal{L}}\mid_{\bbk_2=\bbk_2^*}&=\bzero \label{proof_KKT_3}\\
{\rm tr}(\bbk_1^*\bbm_1)&=0\label{proof_KKT_4}\\
{\rm tr}(\bbk_2^*\bbm_2)&=0\label{proof_KKT_5}\\
{\rm tr}((\bbs-\bbk_1^*-\bbk_2^*)\bbm_S)&=0\label{proof_KKT_6}\\
\lambda_j (R_{0j}^{\rm AL}(\bbk_1^*,\bbk^*_2)-a^*)&=0,\quad
j=1,2\label{proof_KKT_7}
\end{align}
The first KKT condition in (\ref{proof_KKT_1}) implies
$\lambda_1+\lambda_2=1$. We define $\lambda=\lambda_1$ and
consequently $\bar{\lambda}=1-\lambda=\lambda_2$. The second KKT
condition in (\ref{proof_KKT_2}) implies
\begin{align}
\mu_1(\bbk_1^*+\bbk_2^*+\bbsigma_1)^{-1}+\bbm_1&=\mu_0
\lambda(\bbk_1^*+\bbk_2^*+\bbsigma_1)^{-1} +(\mu_0
\bar{\lambda}+\mu_1)(\bbk_1^*+\bbk_2^*+\bbsigma_2)^{-1}+\bbm_S
\label{proof_KKT_2_implies}
\end{align}
Adding $\mu_2(\bbk_1^*+\bbk_2^*+\bbsigma_1)^{-1}$ to both sides
yields (\ref{KKT_1}). Subtracting (\ref{proof_KKT_2}) from
(\ref{proof_KKT_3}) yields (\ref{KKT_2}). Since ${\rm
tr}(\bba\bbb)={\rm tr}(\bbb \bba)$ and ${\rm tr}(\bba \bbb) \geq
0$ for $\bba\succeq \bzero,\bbb\succeq \bzero $,
(\ref{proof_KKT_4})-(\ref{proof_KKT_6}) imply
(\ref{KKT_3})-(\ref{KKT_5}). Furthermore, (\ref{proof_KKT_7})
states the conditions if $R_{01}^{\rm
AL}(\bbk_1^*,\bbk_2^*)>R_{02}^{\rm AL}(\bbk_1^*,\bbk_2^*)$,
$\lambda=0$, if $R_{01}^{\rm AL}(\bbk_1^*,\bbk_2^*)<R_{02}^{\rm
AL}(\bbk_1^*,\bbk_2^*)$, $\lambda=1$, and if $R_{01}^{\rm
AL}(\bbk_1^*,\bbk_2^*)=R_{02}^{\rm AL}(\bbk_1^*,\bbk_2^*)$,
$\lambda$ is arbitrary, i.e., $0<\lambda<1$.

\section{Proof of Lemma~\ref{lemma_enhancement}}
\label{proof_of_lemma_enhancement} To prove the first statement of
the lemma, we note
\begin{align}
(\mu_1+\mu_2)(\bbk_2^*+\tilde{\bbsigma})^{-1}&=(\mu_1+\mu_2)(\bbk_2^*+\bbsigma_2)^{-1}+\bbm_2\label{def_again}
\\
(\mu_1+\mu_2)(\bbk_2^*+\tilde{\bbsigma})^{-1}&=(\mu_1+\mu_2)(\bbk_2^*+\bbsigma_1)^{-1}+\bbm_1
\label{def_implies}
\end{align}
where (\ref{def_again}) is the definition of the new noise
covariance matrix in (\ref{def_enhanced_noise}) and
(\ref{def_implies}) comes from plugging (\ref{def_enhanced_noise})
in (\ref{KKT_2}). Using the fact that for $\bba\succ \bzero$,
$\bbb\succ \bzero$, if $\bba \preceq \bbb$, then $\bba^{-1}\succeq
\bbb^{-1}$ in (\ref{def_again})-(\ref{def_implies}) yields the
first statement of the lemma.

We next show the second statement of the lemma as follows
\begin{align}
\lefteqn{\bbk_1^*+\bbk_2^*+\tilde{\bbsigma}=\bbk_1^*+\left[(\bbk_2^*+\bbsigma_1)^{-1}+\frac{1}{\mu_1+\mu_2}\bbm_1\right]^{-1}
}\label{def_implies_implies}\\
&=\bbk_1^*+\left[\bbi+\frac{1}{\mu_1+\mu_2}(\bbk_2^*+\bbsigma_1)\bbm_1\right]^{-1}
(\bbk_2^*+\bbsigma_1)\\
&=\bbk_1^*+\left[\bbi+\frac{1}{\mu_1+\mu_2}(\bbk_1^*+\bbk_2^*+\bbsigma_1)\bbm_1\right]^{-1}
(\bbk_2^*+\bbsigma_1)
\label{KKT_3_implies}\\
&=\bbk_1^*+\left[(\bbk_1^*+\bbk_2^*+\bbsigma_1)^{-1}+\frac{1}{\mu_1+\mu_2}\bbm_1\right]^{-1}(\bbk_1^*+\bbk_2^*+\bbsigma_1)^{-1}
(\bbk_2^*+\bbsigma_1) \\
&=\bbk_1^*+\left[(\bbk_1^*+\bbk_2^*+\bbsigma_1)^{-1}+\frac{1}{\mu_1+\mu_2}\bbm_1\right]^{-1}(\bbk_1^*+\bbk_2^*+\bbsigma_1)^{-1}
(\bbk_1^*+\bbk_2^*+\bbsigma_1-\bbk_1^*) \\
&=\bbk_1^*+\left[(\bbk_1^*+\bbk_2^*+\bbsigma_1)^{-1}+\frac{1}{\mu_1+\mu_2}\bbm_1\right]^{-1}\nonumber\\
&\quad
-\left[(\bbk_1^*+\bbk_2^*+\bbsigma_1)^{-1}+\frac{1}{\mu_1+\mu_2}\bbm_1\right]^{-1}(\bbk_1^*+\bbk_2^*+\bbsigma_1)^{-1}\bbk_1^*
\\
&=\bbk_1^*+\left[(\bbk_1^*+\bbk_2^*+\bbsigma_1)^{-1}+\frac{1}{\mu_1+\mu_2}\bbm_1\right]^{-1}\nonumber\\
&\quad
-\left[(\bbk_1^*+\bbk_2^*+\bbsigma_1)^{-1}+\frac{1}{\mu_1+\mu_2}\bbm_1\right]^{-1}\left[(\bbk_1^*+\bbk_2^*+\bbsigma_1)^{-1}+\frac{1}{\mu_1+\mu_2}\bbm_1\right]\bbk_1^*
\label{KKT_3_implies_1}\\
&=\left[(\bbk_1^*+\bbk_2^*+\bbsigma_1)^{-1}+\frac{1}{\mu_1+\mu_2}\bbm_1\right]^{-1}
\end{align}
where (\ref{def_implies_implies}) is due to (\ref{def_implies}),
(\ref{KKT_3_implies}) and (\ref{KKT_3_implies_1}) come from
(\ref{KKT_3}).

We now show the third statement of the lemma as follows
\begin{align}
(\bbk_2^*+\tilde{\bbsigma})^{-1}\tilde{\bbsigma}&=\bbi-(\bbk_2^*+\tilde{\bbsigma})^{-1}\bbk_2^*\\
&=\bbi-\left[(\bbk_2^*+\bbsigma_2)^{-1}+\frac{1}{\mu_1+\mu_2}\bbm_2\right]\bbk_2^*
\label{def_implies_again} \\
&=\bbi-(\bbk_2^*+\bbsigma_2)^{-1}\bbk_2^*\label{KKT_4_implies}
\\
&=(\bbk_2^*+\bbsigma_2)^{-1}\bbsigma_2
\end{align}
where (\ref{def_implies_again}) comes from (\ref{def_again}), and
(\ref{KKT_4_implies}) is due to (\ref{KKT_4}).

We finally show the last, i.e., fourth, statement of the lemma as
follows
\begin{align}
(\bbk_1^*+\bbk_2^*+\tilde{\bbsigma})^{-1}(\bbk_2^*+\tilde{\bbsigma})&=
\bbi-(\bbk_1^*+\bbk_2^*+\tilde{\bbsigma})^{-1}\bbk_1^* \\
&=
\bbi-\left[(\bbk_1^*+\bbk_2^*+\bbsigma_1)^{-1}+\frac{1}{\mu_1+\mu_2}\bbm_1\right]\bbk_1^*
\label{third_statement} \\
&= \bbi-(\bbk_1^*+\bbk_2^*+\bbsigma_1)^{-1}\bbk_1^*
\label{KKT_3_implies_2}\\
&=(\bbk_1^*+\bbk_2^*+\bbsigma_1)^{-1}(\bbk_2^*+\bbsigma_1)
\end{align}
where (\ref{third_statement}) comes from the second statement of
this lemma, and (\ref{KKT_3_implies_2}) is due to (\ref{KKT_3}).

\section{Proof of Lemma~\ref{lemma_outer_bound}}
\label{proof_of_lemma_outer_bound}

We prove this lemma for a discrete memoryless broadcast channel
with a transition probability
$p(\tilde{y}_1,\tilde{y}_2,y_1,y_2|x)$ which satisfies
$p(\tilde{y}_1|x)=p(\tilde{y}_2|x)=p(\tilde{y}|x)$ and
\begin{align}
X\rightarrow \tilde{Y}\rightarrow(Y_1,Y_2)
\label{degradedness_cond}
\end{align}
Consequently, Lemma~\ref{lemma_outer_bound} can be concluded from
the proof for this discrete memoryless broadcast channel. We note
that if $(R_0,R_1,R_2)$ is achievable, we need to have
$\epsilon_n,\gamma_n$ such that both $\epsilon_n$ and $\gamma_n$
vanish as $n\rightarrow \infty$, and
\begin{align}
H(W_0|Y_j^n)&\leq n\epsilon_n,\quad j=1,2 \label{Fano_1} \\
H(W_j|\tilde{Y}^n,W_0)&\leq n\epsilon_n,\quad j=1,2\label{Fano_2} \\
I(W_1;Y_2^n,W_0)&\leq n\gamma_n \label{perfect_secrecy_1}\\
I(W_2;Y_1^n,W_0)&\leq n\gamma_n \label{perfect_secrecy_2}
\end{align}
where (\ref{Fano_1})-(\ref{Fano_2}) are due to Fano's lemma, and
(\ref{perfect_secrecy_1})-(\ref{perfect_secrecy_2}) comes from the
perfect secrecy conditions in (\ref{perfect_secrecy_conditions}).
We define the following auxiliary random variables
\begin{align}
U_i&=W_0 \tilde{Y}^{i-1}, \quad i=1,\ldots,n
\end{align}
which satisfy the following Markov chains for all $i,$
\begin{align}
U_i \rightarrow X_i \rightarrow \tilde{Y}_i \rightarrow
(Y_{1i},Y_{2i}) \label{auxiliary_mc}
\end{align}
since the channel is memoryless, and degraded, i.e., satisfies the
Markov chain in (\ref{degradedness_cond}).

We first bound the common message rate $R_0$ as follows
\begin{align}
nR_0&=H(W_0)\\
&\leq I(W_0;Y_1^n)+n\epsilon_n \\
&=\sum_{i=1}^n I(W_0;Y_{1i}|Y_1^{i-1})+n\epsilon_n \\
&\leq \sum_{i=1}^n
I(W_0,\tilde{Y}^{i-1},Y_1^{i-1};Y_{1i})+n\epsilon_n \\
&=\sum_{i=1}^n I(W_0,\tilde{Y}^{i-1};Y_{1i})+n\epsilon_n
\label{degradedness_cond_imply_1} \\
&= \sum_{i=1}^n I(U_i;Y_{1i})+n\epsilon_n \label{common_message_1}
\end{align}
where (\ref{degradedness_cond_imply_1}) comes from the Markov
chain
\begin{align}
Y_1^{i-1}\rightarrow \tilde{Y}^{i-1} \rightarrow (W_0,Y_{1i})
\end{align}
which is a consequence of the fact that the channel is degraded,
i.e., satisfies the Markov chain in (\ref{degradedness_cond}).
Similarly, we can get
\begin{align}
nR_0&\leq \sum_{i=1}^n I(U_i;Y_{2i})+n\epsilon_n
\label{common_message_2}
\end{align}

We next bound the confidential message rate of the enhanced first
user, i.e., $R_1$, as follows
\begin{align}
nR_1&=H(W_1|W_0) \\
&\leq
I(W_1;\tilde{Y}^n|W_0)-I(W_1;Y_2^n|W_0)+n(\epsilon_n+\gamma_n)\\
&\leq I(W_1;\tilde{Y}^n|W_0,Y_2^n)+n(\epsilon_n+\gamma_n)\\
&=\sum_{i=1}^n
I(W_1;\tilde{Y}_i|W_0,Y_2^n,\tilde{Y}^{i-1})+n(\epsilon_n+\gamma_n)\\
&=\sum_{i=1}^n
I(W_1;\tilde{Y}_i|W_0,Y_{2i}^n,\tilde{Y}^{i-1})+n(\epsilon_n+\gamma_n)
\label{degradedness_cond_imply_2} \\
&\leq \sum_{i=1}^n
I(W_1,X_i;\tilde{Y}_i|W_0,Y_{2i}^n,\tilde{Y}^{i-1})+n(\epsilon_n+\gamma_n)\\
&= \sum_{i=1}^n
I(X_i;\tilde{Y}_i|W_0,Y_{2i}^n,\tilde{Y}^{i-1})+n(\epsilon_n+\gamma_n)\label{degradedness_cond_imply_3}\\
&= \sum_{i=1}^n
H(\tilde{Y}_i|W_0,Y_{2i}^n,\tilde{Y}^{i-1})-H(\tilde{Y}_i|W_0,Y_{2i}^n,\tilde{Y}^{i-1},X_i)+n(\epsilon_n+\gamma_n)\\
&\leq \sum_{i=1}^n
H(\tilde{Y}_i|W_0,Y_{2i},\tilde{Y}^{i-1})-H(\tilde{Y}_i|W_0,Y_{2i}^n,\tilde{Y}^{i-1},X_i)+n(\epsilon_n+\gamma_n)\label{conditioning_cannot}\\
&= \sum_{i=1}^n
H(\tilde{Y}_i|W_0,Y_{2i},\tilde{Y}^{i-1})-H(\tilde{Y}_i|W_0,Y_{2i},\tilde{Y}^{i-1},X_i)+n(\epsilon_n+\gamma_n)
\label{memoryless} \\
&=\sum_{i=1}^n I(X_i;\tilde{Y}_i|U_i,Y_{2i})+n(\epsilon_n+\gamma_n) \\
&=\sum_{i=1}^n
I(X_i;\tilde{Y}_i|U_i)-I(X_i;Y_{2i}|U_i)+n(\epsilon_n+\gamma_n)
\label{confidential_message_1}
\end{align}
where (\ref{degradedness_cond_imply_2}) comes from the Markov
chain
\begin{align}
W_0,W_1,Y_{2i}^n \rightarrow \tilde{Y}^{i-1} \rightarrow Y_2^{i-1}
\end{align}
which is a consequence of the fact that the channel is degraded,
i.e., satisfies the Markov chain in (\ref{degradedness_cond}),
(\ref{degradedness_cond_imply_3}) comes from the Markov chain
\begin{align}
W_0,W_1,\tilde{Y}^{i-1},Y_{2(i+1)}^n \rightarrow X_i \rightarrow
\tilde{Y}_i,Y_{2i} \label{memoryless_1}
\end{align}
which is due to the fact that the channel is memoryless,
(\ref{conditioning_cannot}) comes from the fact that conditioning
cannot increase entropy, (\ref{memoryless}) results from the
Markov chain in (\ref{memoryless_1}), and
(\ref{confidential_message_1}) stems from the Markov chain in
(\ref{auxiliary_mc}). Similarly, we can get the following bound on
the confidential message rate of the enhanced second user $R_2$
\begin{align}
nR_2 &\leq \sum_{i=1}^n
I(X_i;\tilde{Y}_i|U_i)-I(X_i;Y_{2,i}|U_i)+n(\epsilon_n+\gamma_n)
\label{confidential_message_2}
\end{align}

The bounds in (\ref{common_message_1}), (\ref{common_message_2}),
(\ref{confidential_message_1}) and (\ref{confidential_message_2})
can be single-letterized yielding the following bounds
\begin{align}
R_0&\leq \min\{I(U;Y_1),I(U;Y_2)\}\\
R_1&\leq I(X;\tilde{Y}|U)-I(X;Y_2|U)\\
R_2&\leq I(X;\tilde{Y}|U)-I(X;Y_1|U)
\end{align}
from which, Lemma~\ref{lemma_outer_bound} can be concluded.

\bibliographystyle{unsrt}
\bibliography{IEEEabrv,references2}

\begin{thebibliography}{10}

\bibitem{Hassibi}
F.~Oggier and B.~Hassibi.
\newblock The secrecy capacity of the {MIMO} wiretap channel.
\newblock Submitted to \emph{IEEE Trans. Inf. Theory}, Oct. 2007. Also
  available at [arXiv:0710.1920].

\bibitem{Wornell}
A.~Khisti and G.~Wornell.
\newblock Secure transmission with multiple antennas {II}: The {MIMOME}
  channel.
\newblock \emph{IEEE Trans. Inf. Theory}, to appear. Also available at
  http://allegro.mit.edu/bin/pubs-search.php.

\bibitem{Ulukus}
S.~Shafiee, N.~Liu, and S.~Ulukus.
\newblock Towards the secrecy capacity of the {G}aussian {MIMO} wire-tap
  channel: {T}he 2-2-1 channel.
\newblock {\em {IEEE} Trans. Inf. Theory}, 55(9):4033--4039, Sep. 2009.

\bibitem{Liu_Common_Confidential}
H.~D. Ly, T.~Liu, and Y.~Liang.
\newblock Multiple-input multiple-output {G}aussian broadcast channels with
  common and confidential messages.
\newblock Submitted to {\it IEEE Trans. Inf. Theory}, Jul. 2009. Also available
  at http://www.ece.tamu.edu/$\sim$tieliu/publications.html.

\bibitem{Ruoheng_MIMO_BC}
R.~Liu, T.~Liu, H.~V. Poor, and S.~Shamai (Shitz).
\newblock Multiple-input multiple-output {G}aussian broadcast channels with
  confidential messages.
\newblock Submitted to \emph{IEEE Trans. Inf. Theory}, Mar. 2009. Also
  available at [arXiv:0903.3786].

\bibitem{Shamai_MIMO}
H.~Weingarten, Y.~Steinberg, and S.~Shamai (Shitz).
\newblock The capacity region of the {G}aussian multiple-input multiple-output
  broadcast channel.
\newblock {\em {IEEE} Trans. Inf. Theory}, 52(9):3936--3964, Sep. 2006.

\bibitem{Liu_Compound}
H.~Weingarten, T.~Liu, , S.~Shamai (Shitz), Y.~Steinberg, and P.~Viswanath.
\newblock The capacity region of the degraded multiple-input multiple-output
  compound broadcast channel.
\newblock {\em {IEEE} Trans. Inf. Theory}, 55(11):5011--5023, Nov. 2009.

\bibitem{Hannan_Common}
H.~Weingarten, Y.~Steinberg, and S.~Shamai (Shitz).
\newblock On the capacity region of the multi-antenna broadcast channel with
  common messages.
\newblock In {\em IEEE ISIT}, Jul. 2006.

\bibitem{hannan_thesis}
H.~Weingarten.
\newblock {\em Multiple-input multiple-output broadcast systems}.
\newblock PhD thesis, Technion, Haifa, Israel, 2007.

\bibitem{Goldsmith_common}
N.~Jindal and A.~Goldsmith.
\newblock Optimal power allocation for parallel broadcast channels with
  independent and common information.
\newblock In {\em IEEE Intl. Symp. Inf. Theory}, page 215, Jun. 2004.

\bibitem{Wyner}
A.~Wyner.
\newblock The wire-tap channel.
\newblock {\em Bell System Technical Journal}, 54(8):1355--1387, Jan. 1975.

\bibitem{Korner}
I.~Csiszar and J.~Korner.
\newblock Broadcast channels with confidential messages.
\newblock {\em {IEEE} Trans. Inf. Theory}, IT-24(3):339--348, May 1978.

\bibitem{Wei_Yu}
W.~Yu and J.~Cioffi.
\newblock Sum capacity of {G}aussian vector broadcast channels.
\newblock {\em {IEEE} Trans. Inf. Theory}, 50(9):1875--1892, Sep. 2004.

\bibitem{MIMO_BC_Secrecy}
E.~Ekrem and S.~Ulukus.
\newblock The secrecy capacity region of the {G}aussian {MIMO} multi-receiver
  wiretap channel.
\newblock Submitted to \emph{IEEE Trans. Inf. Theory}, Mar. 2009. Also
  available at [arXiv:0903.3096].

\bibitem{Biao_Chen}
J.~Xu, Y.~Cao, and B.~Chen.
\newblock Capacity bounds for broadcast channels with confidential messages.
\newblock {\em {IEEE} Trans. Inf. Theory}, 55(10):4529--4542, Oct. 2009.

\bibitem{Diggavi_Cover}
S.~H. Diggavi and T.~M. Cover.
\newblock The worst additive noise constraint under a covariance constraint.
\newblock {\em {IEEE} Trans. Inf. Theory}, 47(7):3072--3081, Nov. 2001.

\bibitem{Ihara}
S.~Ihara.
\newblock On the capacity of channels with additive non-{G}aussian noise.
\newblock {\em Information and Control}, 37(1):34--39, Apr. 1978.

\bibitem{Yingbin_Compound}
Y.~Liang, G.~Kramer, H.~V. Poor, and S.~Shamai (Shitz).
\newblock Compound wire-tap channels.
\newblock {\em EURASIP Journal on Wireless Communications and Networking,
  Special Issue on Wireless Physical Layer Security}, 2009(142374), 2009.

\end{thebibliography}
\end{document}